# The AI Revolution in Education: Will AI Replace or Assist Teachers in Higher Education?

Cecilia Ka Yuk Chan*[1], Louisa H.Y. Tsi[2]

Affiliation: The University of Hong Kong

Address: Centre for the Enhancement of Teaching and Learning (CETL), Room CPD-1.81, Centennial Campus, The University of Hong Kong, Pokfulam, Hong Kong

**Email:** Cecilia.Chan@cetl.hku.hk[1] *Corresponding Author

**Email:** loui0626@hku.hk[2]

**Abstract**

This paper explores the potential of artificial intelligence (AI) in higher education, specifically its capacity to replace or assist human teachers. By reviewing relevant literature and analysing survey data from students and teachers, the study provides a comprehensive perspective on the future role of educators in the face of advancing AI technologies. Findings suggest that although some believe AI may eventually replace teachers, the majority of participants argue that human teachers possess unique qualities, such as critical thinking, creativity, and emotions, which make them irreplaceable. The study also emphasizes the importance of social-emotional competencies developed through human interactions, which AI technologies cannot currently replicate.

The research proposes that teachers can effectively integrate AI to enhance teaching and learning without viewing it as a replacement. To do so, teachers need to understand how AI can work well with teachers and students while avoiding potential pitfalls, develop AI literacy, and address practical issues such as data protection, ethics, and privacy. The study reveals that students value and respect human teachers, even as AI becomes more prevalent in education.

The study also introduces a roadmap for students, teachers, and universities. This roadmap serves as a valuable guide for refining teaching skills, fostering personal connections, and designing curriculums that effectively balance the strengths of human educators with AI technologies.

The future of education lies in the synergy between human teachers and AI. By understanding and refining their unique qualities, teachers, students, and universities can effectively navigate the integration of AI, ensuring a well-rounded and impactful learning experience.

**Introduction**

<p align="center">Robots will replace human teachers by 2027</p>

this was the prediction made by Sir Anthony Seldon in September 2017 (Houser, 2017). With the recent launch of ChatGPT (OpenAI, 2023), a generative AI software that can generate human-like responses to a wide range of topics and the increasing capabilities of AI technologies, the question of whether AI can completely replace the role of teachers is becoming more pressing. It seems that this possibility may be closer than ever before. With the anticipation that more than five million jobs will be replaced by AI, news media is definitely mulling with the idea whether teachers will be the next to be replaced (Cerullo, 2023).

This study aims to explore the perceptions and experiences of students and teachers towards generative AI, seeking to understand whether they believe that AI will replace teachers or whether they will work alongside them.

**AI in Education**

Application of AI in education dates back to the 1950s with the introduction of computer-assisted instruction. Over the decades, it has evolved into intelligent tutoring systems (ITS), which are now widely used for teaching and learning (Nwana, 1990). Currently, a broad range of AI technologies, from ITS offering 1-on-1 tutoring to virtual teaching assistants, is being employed in education. In their review of AI research in education, Goksel & Bozkurt (2019) identified three primary themes that have garnered significant attention from researchers: adaptive learning, expert systems and ITS, and the future role of AI in educational processes. These themes also highlight the prevailing trends in AI development within the field of education.

Studies on AI implementation in education have explored various forms of AI usage in classroom settings, emphasizing the benefits they offer for student learning. Beyond ITS, dialogue-based tutoring systems driven by natural language processing can be utilized to facilitate knowledge co-creation as students engage in conversations with AI (UNESCO, 2021). Web-based AI systems can also provide personalized feedback and handle administrative tasks previously managed by human teachers (Chen, Chen, & Lin, 2020). Chen, Xie, Zou, & Hwang, (2020) highlighted natural language processing and machine learning such as AI-powered language learning platforms as the most commonly adopted AI methods in education due to their effectiveness. As AI technologies continue to advance, they hold promising potential for personalized and adaptive learning, real-time feedback, and intelligent administrative and support systems (Renz, Krishnaraja, & Gronau, 2020), liberating teachers from time-consuming tasks, allowing them to concentrate on higher-level responsibilities like curriculum development and student mentoring.

In addition to traditional computer-based AI systems, innovative technologies such as humanoid robots, chatbots, and virtual reality systems are being integrated into the educational process (Chen, Chen & Lin, 2020; ThinkML Team, 2022; UNESCO, 2021). These technologies can enhance student engagement by providing interactive, personalized, and immersive learning environments (Malik, Tayal & Vij, 2019; Chen, Chen & Lin, 2020). And not just that, in a study by Blikstein (2016), he found that AI-supported classrooms yielded higher engagement levels and greater student achievement compared to traditional classrooms. Consequently, research into the integration of AI technologies in education is expected to accelerate, as the potential benefits of AI in education become more widely recognized.

**Literature on AI vs. Teachers**

Amid the growing development and implementation of AI in education, concerns have emerged regarding the potential for AI to replace teachers altogether. Some argue that AI is better equipped than human educators to deliver standardized content and assessments, and can work tirelessly without fatigue or bias. However, others contend that AI lacks the empathy and emotional intelligence necessary for effective teaching and learning.

On the bright side, the wide array of functions that AI is capable of performing can take over some of the duties that teachers are responsible for. Teachers need to allocate a certain amount of time in handling administrative tasks such as attendance checking, assignment and classroom monitoring, and paperwork. With the introduction of AI, not only can these tasks be relieved from teachers, but also be accomplished in a much more efficient way. Multiple studies and reports have provided evidence to support the notion that the time-consuming administrative tasks involved in the teaching and learning process can be done through AI technologies while not compromising the tasks quality (Chen, Chen & Lin, 2020; Felix, 2020; UNESCO, 2021). A survey revealed that teachers spend as much as 15% of their time on such tasks (McKinsey & Company, 2020). Utilizing AI technologies for these tasks can save time, allowing teachers to focus on addressing students' learning needs. As pointed out by Popenici & Kerr (2017), AI poses a particular threat to university staff and teaching assistants primarily responsible for administrative duties. Prof. Luckin from UCL Knowledge Lab even predicted that every teacher could have a dedicated AI assistant by 2030 (Luckin, Holmes, Griffiths, & Forcier, 2016). Moreover, AI has the capability to assist teachers in student assessment, as developments in natural language processing facilitate applications such as plagiarism detection, assessment scoring, and automated feedback provision (Chen, Chen & Lin, 2020; Goksel and Bozkurt, 2019). Owing to their dependence on algorithms and data, AI technologies can provide more objective and efficient feedback compared to human teachers (Celik, Dindar, Muukkonen & Järvelä, 2022; Terzopoulos & Satratzemi, 2019). Furthermore, tracking the learning progress of a group of students can be challenging for teachers. However, AI can assist in this area by ensuring more effective monitoring of students' learning progress, as various ITSs include functions to track and record each student's learning journey, enabling teachers to gain a better understanding of their students and intervene when needed (Celik, Dindar, Muukkonen & Järvelä, 2022). In the context of language education, AI technologies have fostered a student-centered approach and increased learner autonomy by allowing students to monitor their own learning pace through AI-supported systems (Pokrivcakova, 2019).

While there are certainly advantages to using AI in education, it's important to recognize that AI has limitations that raise doubts about the feasibility of replacing human teachers with AI. Firstly, AI currently lacks sentience and self-awareness, producing only mechanical responses without emotions (Felix, 2020; Pavlik, 2023). Timms (2016) emphasized that emotional support from teachers is essential for student engagement and motivation, which AI technologies have yet to automate (Schiff, 2020, p.341). Moreover, values and social norms cannot be quantified and reduced to algorithms (Felix, 2020). Thus, humans still outperform AI in social and emotional aspects, emphasizing the irreplaceable role of human teachers (Jarrahi, 2018). Secondly, AI-student interactions fall short of the educational value provided by real-life human interactions. A crucial element in education is how teachers motivate and facilitate students in their learning. As mentioned by Schiff (2020), "a teacher must know their students" in order to deliver effective guidance and facilitation for the students (p.335). Additionally, relying on AI and online platforms may limit peer interactions and hinder the development of essential social skills (Wogu, Misra, Olu-Owolabi, Assibong, & Udoh, 2018). Teacher-student relationships, peer interactions, and connections between students, families, communities, and schools form the "social milieu" of education, where teaching and learning occurs (Yang & Zhang, 2019, p. 4). Despite AI's capabilities, scholars only view AI as "cognitive prostheses" that can aid teaching and learning, but not yet capable of replacing the values of human thoughts or collaborative relationships between teachers and students (Cope, Kalantzis & Searsmith, 2021; Felix, 2020; Kim, Lee & Cho, 2022). Thirdly, other concerns on the limitations and drawbacks of AI technologies also confine their roles in education. Some of the notable concerns include the dubious technical capacity and reliability of algorithms (Celik, Dindar, Muukkonen & Järvelä, 2022), the necessary human input or training from human operators in order for AI to function properly (Wilson & Daugherty, 2018), inequality and prejudice issues arising from reliance on AI (Wogu, Misra, Olu-Owolabi, Assibong, & Udoh, 2018), and the comparative disadvantage of AI in holistic and visionary thinking (Jarrahi, 2018). Overall, Popenici & Kerr (2017) concluded the value of AI at its current state of development lies in augmenting teachers rather than replacing them completely. Based on a review of the literature, there are 8 categories and 26 aspects that highlight the unique skills, qualities, and experiences of human teachers that AI cannot yet replicate, (note that it is possible some aspects may fit into multiple categories). Table 1 shows a roadmap demonstrating the 8 categories and 26 aspects highlighting the limitations of AI in education from the literature.

| Emotional and Interpersonal Skills |
| --- |
| 1. Human connection: The emotional bond and interpersonal skills that teachers have are essential for students' personal growth and development. Teachers can understand, empathize, and motivate students in a way that AI cannot. |
| 2. Cultural sensitivity: Teachers can understand and navigate the cultural nuances of their students and adapt their teaching approach accordingly. AI systems may struggle to replicate this level of cultural sensitivity. |
| 3. Developing resilience and perseverance: Teachers play a vital role in helping students develop resilience and perseverance by offering support, guidance, and encouragement in the face of challenges. AI systems may not be as adept at providing this emotional and motivational support. |
| 4. Building trust and rapport: Teachers build trust and rapport with their students through personal interactions, which helps create a positive learning environment. AI systems may struggle to replicate the trust-building process that human teachers can foster. |
| 5. Social and emotional learning: Teachers support students' social and emotional learning by modeling appropriate behavior, discussing emotions, and helping students develop self-awareness and empathy. AI systems might have limited capacity to engage in these complex aspects of human development. |
| 6. Role model: Teachers serve as role models for their students, demonstrating a passion for learning, a commitment to personal growth, and the importance of hard work and perseverance. AI systems, while helpful in providing information, cannot serve as role models in the same way human teachers can. |
| Pedagogical Skills |
| 1. Real-world context: Teachers can provide real-world examples and experiences that AI systems may not be able to offer, which helps students better understand and relate to the material being taught. |
| 2. Encouraging curiosity: Teachers can inspire students to be curious by sharing their own enthusiasm for learning and fostering a growth mindset. While AI systems can provide resources and support, they may not be able to instill the same sense of curiosity and passion for learning. |
| 3. Professional development: Teachers continuously engage in professional development to improve their teaching practice and stay up-to-date with the latest educational research. AI can provide resources and tools for professional development, but human interaction and discussion with colleagues remain essential for growth and improvement. |
| 4. Encouraging debate and open-mindedness: Teachers can foster an environment of open-mindedness and encourage debate by presenting diverse perspectives, asking challenging questions, and facilitating discussions. AI systems may not have the same level of effectiveness in stimulating meaningful debates and encouraging open-mindedness. |
| 5. Conflict resolution: Teachers often help mediate conflicts between students and teach them essential conflict resolution skills. AI systems may not be as effective in addressing conflicts and facilitating resolution. |
| 6. Experiential learning: Teachers often design and facilitate hands-on, experiential learning opportunities for their students, such as field trips, lab work, or other interactive experiences. While AI systems can support some aspects of experiential learning, human teachers remain essential for planning, executing, and supervising these activities. |
| Holistic Competency Development |
| 1. Adaptability: Teachers are able to adjust their teaching methods and strategies based on the specific needs of their students, which AI systems may struggle to do as effectively. |

| | |
|---|---|
| 2. | Critical thinking and creativity: Teachers can foster creativity and critical thinking in students by designing engaging lessons, asking thought-provoking questions, and encouraging open discussions. AI systems are limited in their ability to engage in these activities. |
| 3. | Collaboration and teamwork: Teachers help students develop collaboration and teamwork skills through group projects, discussions, and other cooperative activities. AI systems can facilitate some collaborative tasks, but the human touch is still essential for promoting a true sense of teamwork. |
| 4. | Nurturing creativity and innovation: Teachers help cultivate creativity and innovation by allowing students to explore new ideas, take risks, and think outside the box. AI systems can offer some support for creative tasks, but they may lack the ability to truly nurture and develop creative potential in students |
| 5. | Teaching life skills: Teachers often help students develop essential life skills, such as time management, goal setting, and decision making. AI systems can provide resources and tools for teaching these skills, but the guidance and personal experiences shared by human teachers can be invaluable for students. |
| **Ethical and Moral Considerations** | |
| 1. | Ethical considerations: There are numerous ethical concerns surrounding the use of AI in education, such as data privacy, algorithmic bias, and the potential for misuse of AI-generated content. |
| 2. | Moral and ethical guidance: Teachers often play a role in shaping students' moral and ethical values by discussing complex issues and encouraging reflection. AI systems lack the capability to provide such guidance. |
| **Personalized Support** | |
| 1. | Behavior management: Teachers are skilled at managing classroom behavior and addressing issues as they arise. AI systems might not be as effective in dealing with behavioral issues or understanding the underlying reasons behind them. |
| 2. | Support for students with special needs: Teachers provide tailored support to students with special needs, accommodating their learning styles and addressing any unique challenges they may face. AI systems can offer some level of personalization, but the nuanced understanding and empathy of human teachers are critical for effectively supporting students with special needs. |
| **Community and Civic Engagement** | |
| 1. | Parent-teacher communication: Teachers communicate with parents to discuss students' progress, share concerns, and offer guidance. AI systems could assist with some communication tasks, but the personal touch and emotional understanding that teachers bring are crucial for effective parent-teacher communication. |
| 2. | Encouraging civic engagement: Teachers help students understand the importance of civic engagement and develop a sense of responsibility to their community and society. AI systems may provide information about civic engagement opportunities but may not be as effective in inspiring students to take action and become engaged citizens. |
| **Career and Personal Mentorship** | |
| 1. | Career guidance and mentorship: Teachers can offer valuable career guidance and mentorship by sharing their own experiences, offering advice, and connecting students with relevant resources and opportunities. AI systems may not be able to provide the same level of personal insight and guidance. |
| **Physical and Artistic Education** | |
| 1. | Physical education and sports coaching: Teachers play a crucial role in promoting physical fitness and coaching sports teams. AI systems may assist with tracking performance data or providing some instructional content, but they cannot replace the hands-on guidance and encouragement of human teachers and coaches. |
| 2. | Artistic expression and appreciation: Teachers help students develop their artistic skills and appreciation for various art forms. AI systems can offer some support for creative tasks but may lack the ability to inspire artistic expression and foster a deep appreciation for the arts. |

Table 1: Human Teachers' Unique Qualities: A Roadmap Highlighting the Strengths of Teacher Vs the Limitations of AI in Education

**Literature on Collaboration between AI and Teachers**

Rather than presenting a dichotomy between AI and teachers, many researchers argue that the most effective approach involves collaboration between the two. In corporate settings, such collaboration has been found to be conducive for achieving the most significant improvements in performance (Wilson & Daugherty, 2018). Synergistic relationship between AI and humans in organizational contexts also shed lights on how both can complement each other's weaknesses (Jarrahi, 2018). AI can support teachers by automating routine tasks, providing personalized feedback, and generating insights from student data. Conversely, teachers can offer the human touch that AI lacks by providing emotional support, facilitating social interaction, and adding context to learning experiences.

Siemens and Baker (2012) found that a combination of human and machine intelligence resulted in more effective learning outcomes than working alone. In their review of literature within the field of AI in education, Roll & Wylie (2016) also revealed the increased collaboration of teachers and AI technologies in creating interactive learning environment (ILE) over the past 20 years. Consequently, researchers are increasingly focusing on conceptualizing the relationship between human teachers and AI to enhance both the learning capabilities of AI and teaching in general (Chen, Chen & Lin, 2020).

In 2016, Georgia Tech introduced a virtual teaching assistant named Jill Watson by using the AI function from IBM and it was responsible for engaging in conversation with students on online forums, answering queries concerning the coursework and lesson content (Georgia Tech, 2016). In 2019, the humanoid robotic lecturer Yuki started helping out in lectures in Germany by delivering the content put in by technicians beforehand and performing some other administrative duties (RoboticsBiz, 2019). Technologies like these were expected to be playing an even more important role in aiding teachers in class and providing support to cater the learning needs of students (Popenici & Kerr, 2017; Timms, 2016).

Konjin & Hoorn (2020) demonstrated how AI could be effectively implemented in teaching and learning by introducing social robots capable of offering verbal encouragement and gestures for remedial teaching tasks in mathematics. In language education, conversational AI has been found to provide individualized feedback and practice opportunities for students, which can be lacking due to teacher workload, allowing teachers to focus on designing and decision-making aspects of the instructional process (Ji, Han & Ko, 2023). Furthermore, AI learning ecosystems can be used by teachers for assessment purposes, facilitating peer reviews, generating machine-generated feedback, and more (Cope, Kalantzis, & Searsmith, 2021).

There are also many intelligent functions of AI technologies that teachers can use to improve teaching such as the ITS's ability to record students' learning and characteristics (Schiff, 2020), intelligent tutoring that perceives and analyses the emotions of students (Yang & Zhang, 2019), and sensors, monitors and facial recognition cameras that help teachers manage the class and handle learning tasks (Timms, 2016). In fact, the prevalent application of AI in teaching and learning contributed to the burgeoning of EdTech companies and introduced more related practitioners into the field of education such as instructional designers and course developers who specialize in e-learning and mobile learning (Renz, Krishnaraja, & Gronau, 2020).

**Rationale for this study**

The research questions of the study are

1. whether generative AI technologies can replace teachers, and

2. how generative AI technologies can work with or against teachers.

The rationale for this study can be explained on two levels. First, by understanding the potential impact of AI technologies on the role of human teachers, it assists educators in preparing for the integration of AI into educational settings. As AI continues to develop and evolve at an unprecedented rate, it is crucial for educators to be ready for the changes that will impact traditional teaching and learning. Second, as both the opportunities and challenges of implementing AI in education are being explored, this study can offer new insights on the discussion of how AI and human educators can collaborate to enhance the quality of education. Instead of creating a dichotomy between the two, educators need to adapt and determine the best way to optimize their value in teaching and learning while co-exist and co-partner with the ever-evolving AI technologies.

**Methodology**

In this study, a survey design was employed to collect data on the usage and perceptions of generative AI in teaching and learning from students and teachers in Hong Kong universities. The online questionnaire featured a combination of closed-ended and open-ended questions, addressing topics such as the integration of AI technologies like ChatGPT in higher education, potential risks associated with these technologies, and their impact on teaching and learning.

Participants were recruited through bulk email invitations, and a convenience sampling technique was applied, selecting respondents based on their availability and willingness to participate. Before completing the survey, participants were provided with an informed consent form.

The final sample consisted of 384 undergraduate and postgraduate students and 144 teachers from various disciplines. Descriptive analysis was employed to analyse the survey data, while a thematic analysis approach was used to examine responses from the open-ended questions. Prior to the main survey, two rounds of pilot study were conducted with 20 students and teachers, chosen randomly. The survey was revised based on feedback from the pilot study and discussions with a team of researchers working on the project.

**Findings**

*Quantitative Data Findings*

The findings reveal that both groups provided valuable insights into these two research questions (refer to Table 2). This table presents a survey on students' and teachers' perceptions regarding the potential of AI to replace teachers. The survey contains 11 items, with higher scores indicating greater agreement. The table displays the number of participants (n), mean (M), and standard deviation (SD) for both students and teachers.

From the findings, students were found to have a higher mean score (M=3.86, SD=1.008) than teachers (M=3.61, SD=1.183), suggesting that students are more open to integrating generative AI technologies into their learning practices compared to teachers (t=2.238, df=215.111, p=.026). Students, especially younger generations, are more accustomed to using digital tools and technology for various aspects of their lives. They may be more open to embracing AI technologies for educational purposes and value the convenience and accessibility they provide. In relation to student learning, students (M=3.08, SD=1.142) had a higher mean score than teachers (M=2.67, SD=1.127), suggesting that students believe generative AI technologies can provide guidance for coursework as effectively as human teachers, more so than teachers do (t=3.636, df=531, p=<.001). AI technologies can provide immediate feedback, answers, or guidance without students having to wait for a teacher's availability or response. This enables students to get help with their academic tasks whenever they need it, fostering a sense of independence and control over their learning process. Similarly, students (M=3.47, SD=.979) were found to have a higher mean score than teachers (M=3.29, SD=1.115) in relation to academic performance, suggesting that students are more optimistic about the potential of generative AI technologies to improve their overall academic performance (t=1.792, df=507, p=.074). Also, students (M=3.32, SD=1.162) had a higher mean score than teachers (M=3.01, SD=1.273), indicating that students believe generative AI technologies can help them become better writers more than teachers do (t=2.577, df=525, p=.010).

Students and teachers both believe generative AI can bring them unique insights and perspectives, although, students (M=3.74, SD=1.076) had a slightly higher mean score than teachers (M=3.47, SD=1.079), suggesting that students believe generative AI technologies can provide unique insights and perspectives that they may not have thought of themselves, more so than teachers do (t=2.533, df=526, p=.012). Students enjoy the 24/7 availability of AI technologies (M=4.13, SD=.826) with a higher mean score than teachers (M=3.69, SD=1.068), (t=4.483, df=216.752, p=<.001). Students have diverse learning styles and preferences. Some may prefer to study late at night or during the weekends, while others might find it challenging to focus during conventional school hours. The 24/7 availability of AI technologies caters to these individual preferences, enabling students to access educational resources whenever it suits them best.

With technology on their finger tip, it is not difficult to understand that students can ask generative AI questions that they may not ask their human teachers. In this case, teachers (M=3.73, SD=.883) had a higher mean score than students (M=3.39, SD=1.094), indicating that teachers believe students are more likely to ask questions to generative AI technologies that they would not ask their human teachers (t=-3.695, df=308.968, p=<.001). Students may feel more comfortable asking questions or seeking help from AI technologies due to the anonymity they offer without the fear of judgment or embarrassment. This could be particularly beneficial for students who are shy, introverted, or hesitant to ask questions in a classroom setting. Teachers are more concerned than students about the potential negative impact of AI technologies on the development of generic or transferable skills (t=-3.087, df=523, p=.002), teachers (M=3.48, SD=1.238) had a higher mean score than students (M=3.10, SD=1.227). In relation to AI detection, teachers (M=2.20, SD=1.125) had a lower mean score than students (M=2.54, SD=1.102), indicating that teachers are less confident in their ability to identify the use of generative AI technologies in students' assignments (t=3.066, df=486, p=.002).

Regarding the main focus of the study, whether AI technologies can replace teachers, both students (M=2.02, SD=.919) and teachers (M=2.03, SD=.946) had similar mean scores, indicating that both groups do not strongly believe that AI technologies will replace teachers in the future (t=-.057, df=539, p=.955). This echoed the question on whether students would pursue a degree through a fully online AI-assisted program, students (M=2.70, SD=1.270) and teachers (M=2.84, SD=1.220) had similar mean scores, indicating that both groups are not strongly in favor of pursuing a degree through a fully online AI-assisted program (t=-1.140, df=514, p=.255).

Both students and teachers showed openness to integrating generative AI technologies into teaching and learning, the survey results suggest that students tend to have a slightly more positive outlook towards the integration and potential benefits of generative AI technologies in education as compared to teachers. They recognized the potential benefits of these technologies, such as improving academic performance, assisting in writing, and providing unique insights. However, both groups do not strongly believe that AI technologies will replace teachers in the future. Teachers appear to be more concerned about the potential negative impact of AI on the development of generic or transferable skills. These findings highlight the need for further exploration into how AI and human educators can collaborate effectively to enhance the quality of education, rather than creating a dichotomy between the two.

| Item | Students n | Students M(SD) | Teachers n | Teachers M(SD) | MD | t | (df) | p |
|---|---|---|---|---|---|---|---|---|
| 1. I envision integrating generative AI technologies like ChatGPT into my teaching and learning practices in the future. | 382 | 3.86 (1.008) | 139 | 3.61 (1.183) | .252 | 2.238 | 215.111 | .026 |
| 2. Generative AI technologies such as ChatGPT can provide guidance for coursework as effectively as human teachers. | 389 | 3.08 (1.142) | 144 | 2.67 (1.127) | .404 | 3.636 | 531 | <.001* |
| 3. I believe Generative AI technologies such as ChatGPT can improve my / students' overall academic performance. | 371 | 3.47 (.979) | 138 | 3.29 (1.115) | .182 | 1.792 | 507 | .074 |
| 4. I think generative AI technologies such as ChatGPT can help me / students become a better writer. | 384 | 3.32 (1.162) | 143 | 3.01 (1.273) | .301 | 2.577 | 525 | .010* |
| 5. I believe AI technologies such as ChatGPT can provide me / students with unique insights and perspectives that I / they may not have thought of themselves. | 387 | 3.74 (1.076) | 141 | 3.47 (1.079) | .268 | 2.533 | 526 | .012* |
| 6. I think AI technologies such as ChatGPT is a great tool (for students) as it is available 24/7. | 392 | 4.13 (.826) | 148 | 3.69 (1.068) | .436 | 4.483 | 216.752 | <.001* |
| 7. I / Students can ask questions to generative AI technologies such as ChatGPT that I / they would otherwise not voice out to their teacher. | 387 | 3.39 (1.094) | 142 | 3.73 (.883) | -.342 | -3.695 | 308.968 | <.001* |
| 8. Generative AI technologies such as ChatGPT will hinder my / students' development of generic or transferable skills such as teamwork, problem-solving, and leadership skills. | 385 | 3.10 (1.227) | 140 | 3.48 (1.238) | -.375 | -3.087 | 523 | .002* |
| 9. Teachers can already accurately identify a student's usage of generative AI technologies to partially complete an assignment. | 353 | 2.54 (1.102) | 135 | 2.20 (1.125) | .344 | 3.066 | 486 | .002* |
| 10. AI technologies like ChatGPT will replace teachers in the future. | 397 | 2.02 (.919) | 144 | 2.03 (.946) | -.005 | -.057 | 539 | .955 |
| 11. If a fully online programme with the assistance of a personalized AI tutor was available, I/ students should be open to pursuing their degree through this option. | 379 | 2.70 (1.270) | 137 | 2.84 (1.220) | -.143 | -1.140 | 514 | .255 |

Table 2: Descriptive analysis and T-test results

*Qualitative Data Findings*

**Generative AI technologies replacing teachers**

Concerning whether generative AI technologies can replace teachers, the qualitative findings revealed valuable perceptions from the teachers and students. In general, majority of the teachers and students cannot foresee generative AI replacing teachers. Some students who believe teachers are irreplaceable perceive generative AI technologies as auxiliary tools controlled by human beings. One student asserted, *"It's ultimately a toy to play with by asking questions. An auxiliary tool at most."* Another student stated, *"I don't think I have a great concern because, as for the time being, all these tools are still controlled by humans."*

**(1) Replacing the Role of the Teacher**

Some teachers and students perceive the possibility that generative AI technologies can replace teachers' roles. A teacher stated, *"I'd like to see if AI can teach students like a human. Students can already communicate with AI as a human. If yes, AI can replace teachers."* A student also mentioned, *"Knowledge and data are easily accessible by every stakeholder. Teachers will lose their value and contribution if they continue to use the old way of teaching."*

However, some teachers perceive their roles as irreplaceable. *"It cannot replace teachers' input, at least at the moment, because AI cannot tell the students in detail, and the students need to know what question to ask,"* emphasized a teacher. Another teacher also mentioned, *"The tech is a game-changer - but it's our job to teach students how to work in a changing world."*

**(2) Replacing the Social-Emotional competencies developed from a Teacher's Human Connection**

Some teachers and students perceive teachers as irreplaceable because generative AI technologies cannot replace human thinking, creativity, and emotions. A student commented on generative AI technologies, *"It (AI) can gather and properly process the information, but it cannot make expansion or innovate."* He further emphasized, *"Creativity and emotion are the most precious personality of our human at whatever time."* Apart from cognition, some teachers also perceive generative AI technologies are incapable of mastering cultural qualities and traditional values as humans. A teacher commented, *"I believe such qualities have to be mastered by students/scholars in this field through the accumulation of knowledge and experience, as well as interaction with different people across a range of contexts. These cannot be replaced by AI technologies."*

**Generative AI technologies working with teachers**

Based on the quantitative findings of this study, it is evident that generative AI technologies have the potential to collaborate effectively with teachers, yielding a range of benefits. In the qualitative findings below, teachers and students voiced out how these technologies can facilitate their work and enhance student learning. By leveraging the capabilities of generative AI, educators can optimize their teaching strategies, thereby enriching the overall learning process and outcomes for their students. The symbiotic relationship between teachers and generative AI technologies demonstrates the transformative impact of AI integration in the educational domain.

**(1) Enhancing Course Planning, Design, and Pedagogy**

The study's qualitative findings reveal that some teachers are utilizing AI technologies, particularly generative AI, to enhance course planning, design, and pedagogy. Teachers employ these technologies to brainstorm, summarize ideas, gather information, generate inspiration, and design engaging courses, ultimately improving their teaching methods. The findings show that teachers use generative AI technologies for preliminary searching and identifying knowledge in various areas, allowing for "*a broader consideration of topics*" and "*a quick understanding of ideas to save time*".

In addition, some teachers also collaborate with generative AI technologies to design engaging courses. They share their experiences of using generative AI technologies in course design, such as "*generating scenarios tailored to students' majors, engaging them in problem-solving related to real-world issues, and creating MCQ questions for Kahoot games*", in order to make their courses more interesting and engaging.

Another teacher mentioned instructing students to "*generate an assignment using ChatGPT to answer a prompt and critique that essay and make corrections using ChatGPT*". Some students also suggested teachers to evaluate students' learning outcomes with these technologies. For instance, a student mentioned, "*ChatGPT can provide features such as automatic scoring and speech recognition to help teachers better assess students' learning outcomes and oral presentation skills.*"

**(2) Developing Students' Research and Writing Skills**

One remarkable aspect of generative AI is its proficiency in producing well-crafted text (Morris, 2023), teachers can take advantage of this to enhance students' writing and research skills with the technology. One teacher noted, "*AI can definitely function as a guide to writing. It writes excellent papers in terms of structure, clarity, and logic (at least by appearance)*". Aligning with teachers' perceptions, a student also mentioned, "*ChatGPT could be used to help students improve their writing skills by generating suggestions for improving sentence structure, grammar, and vocabulary*". This aligns with the quantitative findings.

Teachers also work with generative AI technologies to assist students in research and enhance students' research experience by helping students to "*identify keywords for a topic of research and test search phrases for researching*" and "*search references and prepare the reference sections of academic papers.*"

**(3) Preparing Students for an AI-driven Workplace and Future**

Teachers focus on ensuring students become proficient at using AI technologies and understand their implications on the workplace and future careers. A teacher noted, "*Students will need to learn how to use the tool to enhance their productivity in the workplace.*" Another teacher also emphasized, "*This is a tech that my students will have access to all their lives. It is my job to make sure they are proficient at using it.*"

**(4) Improving Time Efficiency and Reducing Costs**

Teachers agree that generative AI technologies improve their time efficiency and reduce costs, the technologies "*speed up routine tasks*", "*fasten the pace of lesson preparation*", and design assessments. One teacher mentioned, it can "*help me finish some administration work,*

*especially regarding course logistics and tutorial registration, as these are very time-consuming and labour-intensive*" and "*generate some email templates*". Students also perceive similar benefits. A student noted, "*For future teaching, I believe that the use of ChatGPT will reduce teachers' workload for answering students' questions. I may also use it to generate some lesson plans.*"

**(5) Encouraging Personalized Learning and Immediate Feedback**

AI technologies are utilized as virtual tutors, providing personalized learning experiences and immediate feedback on student responses. Generative AI technologies such as ChatGPT are capable of acting as a virtual intelligent tutor (Qadir, 2022), some teachers work with generative AI technologies to engage in personalized learning. One teacher commented, "*I think AI like ChatGPT could be used as a personal tutor. But instead of showing the answers directly, providing advice or directions might be better.*" In addition, some teachers also perceive using these technologies to provide students with immediate feedback. For instance, a teacher suggested, "*Perhaps it's possible to have ChatGPT review student writing and give them feedback on how and where to improve their writing*". A student also agree that teachers can use generative AI technologies such as ChatGPT to "*provide immediate feedback on student responses.*"

**Generative AI Technologies Working Against Teachers**

While some teachers and students perceive the possibilities to work with generative AI technologies, some perceive these technologies may work against teachers for specific tasks such as the development of holistic competencies indicated in the quantitative findings.

**(1) Lacking AI Literacy**

Some teachers perceive generative AI technologies will work against teachers if they fail to provide proper guidelines and training to students using these technologies properly. One teacher mentioned, "*It's very important to train our students how to use such AI technologies 'sensibly' with 'a responsible and professional manner'*". Some teachers emphasised the lack of AI literacy could be detrimental. They suggested AI literacy for staff and students, for example, Ethics of use (when/why) equity, privacy/property, knowledge of affordances (features/benefits/limits of various tools), effective use (e.g prompt engineering), critique/evaluation of outputs, role/integration in workflows/product in study and professional settings. A student also raises her concern, "*AI technologies will only become more mature which leads to heightened difficulties in identifying works of man vs. AI technologies.*" In light of this, a student commented, "*It is very important for tutors to figure out whether a student is using AI*" and how AI can be used to enhance learning.

**(2) Failing to Ensure Equity, Preventing Academic Misconduct and Addressing Governance of Generative AI Technologies**

Some teachers raise the concerns that generative AI technologies will work against teachers if students violate ethical and academic integrity issues when using these technologies. A teacher commented, "*Seeing as the technology is based on rehashing existing data, including original work by unacknowledged (human) creators, AI is by definition a plagiarizing technology.*" This may further lead to a loss of trust between students and teachers. "*The widespread use of AI will definitely make professor lose trust in students*," stated a student. Therefore, in order to ensure proper and fair use of generative AI technologies, it is necessity of establishing regulations and frameworks. A teacher noted, "*I think it is a boon to teaching and learning. However, some regularity mechanism must be undertaken for fair usage in the university for the students.*"

In addition, teachers also emphasized the importance of avoiding students over-relying on these technologies. A teacher commented on the consequences of students' over-reliance on generative AI technologies, "*There will be a dearth of original ideas as people become lazy and use AI.*" Another teacher also expressed his fear of students' over-relying on these technologies, "*I fear this is fostering over-reliance on tools like this, limiting the development of critical thinking.*"

**(3) Undermining Holistic Competency Development**

Some teachers express concerns that generative AI technologies may hinder students' potential for new and original discoveries. As one teacher stated, "*I require students to discover things for themselves and to do their own research*", emphasizing the counterproductive nature of generative AI technologies that only "copy existing information". Other teachers believe that generative AI technologies undermine students' learning by producing text with little or incomplete understanding of the content, as it no longer demonstrates learning. Furthermore, teachers acknowledge the failure of generative AI technologies to ensure the accuracy of AI-generated content, potentially leading to factuality errors, which could undermine student learning. While generative AI technologies can assist in idea generation and research, some teachers believe that they may hinder the development of students' cognitive and holistic competencies. These teachers believe that generative AI technologies may bring "*negative impacts on creativity*", "*prevent students from critical thinking*" and "*handicap students' intellectual and mental development*". However, while generative AI technologies may not be able to replace the development of these competencies, some teachers believe that they may be able to use the technologies to develop them such as "*practicing exercising judgment*".

**Discussion**

The study presents a complex and nuanced perspective on teacher's potential role in the future of education as generative AI technologies immersed in education. Some teachers and students believe that these technologies may eventually replace teachers, as evidenced by a teacher's statement expressing the desire to see if AI can teach students like a human, and a student's concern about teachers losing their value if they continue using traditional teaching methods. However, most participants argue that teachers' roles are irreplaceable due to the unique human qualities they bring to the educational process, such as critical thinking, creativity, and emotions. One student emphasized that creativity and emotion remain the most precious aspects of human personality, which AI cannot replicate or replace. This is also evidenced in the quantitative data with teachers showing a slightly higher concern on students' development of holistic competency. Teachers are responsible for providing a well-rounded education that not only focuses on subject-specific knowledge but also nurtures students' overall development (Chan, Fong, Luk, & Ho, 2017). They may be concerned that an overreliance on AI technologies could lead to a more fragmented learning experience, where students might excel in acquiring knowledge but struggle to develop essential life skills. In addition, they might be more aware of the importance of developing holistic competencies, such as problem-solving, critical thinking, communication, and teamwork, for students' long-term success in their personal and professional lives.

Additionally, the study found that some teachers believe generative AI technologies cannot replace the social-emotional competencies developed through interactions with human educators. These participants argue that, beyond cognitive abilities, AI technologies lack the capacity to master cultural qualities and traditional values in the same way humans can through knowledge accumulation, experience, and interaction with diverse individuals across various contexts. Furthermore, teachers serve as crucial communicators with parents and the community while inspiring civic engagement and providing career guidance and mentorship. Human teachers are indispensable in promoting physical fitness and artistic expression, nurturing students' appreciation for the arts, and designing hands-on, experiential learning opportunities. Despite AI's ability to provide information and support, it lacks the emotional intelligence, cultural sensitivity, and capacity for trust-building essential for students' personal growth and development. Human teachers excel in their ability to adapt their teaching methods and strategies to individual students' needs and engage them in critical thinking, creativity, and collaboration. They also play a vital role in guiding students through moral and ethical dilemmas, managing classroom behavior, and addressing the unique challenges faced by students with special needs. These aspects of teaching illustrate the irreplaceable value of human teachers in education, even as AI continues to advance and support the learning process. These findings align with the literature as shown Table 1.

The findings also suggest that, instead of considering generative AI technologies as tools to replace teachers, teachers can incorporate these technologies to enhance teaching and learning. However, to incorporate these tools effectively, teachers should have a comprehensive understanding of the dimensions where generative AI technologies can work well with teachers and students, the conditions that need to be avoided to prevent generative AI technologies from working against teachers. This is also supported by scholars, who believe that it will be crucial to develop teachers' competencies and capacities (AI literacy) for collaborating with AI to support their teaching in the future (Kim, Lee & Cho, 2022). Other practical issues such as data protection, ethics, and privacy must also be resolved before further integrating AI into the classroom (Renz, Krishnaraja, & Gronau, 2020).

Contrary to the belief that young generation might prefer technology approach to learning, this study's findings reveal that teachers are valued and possess many qualities that students respect and appreciate. Generative AI is here to stay, and it will help us in our personal, social, and professional lives. To become more efficient, teachers and students need to harness AI, ensuring that guidelines and training are available to upgrade their AI literacy and collaborate with AI effectively and prepare students' future. From the findings, neither students nor teachers foresee a future without teachers in the classroom. However, if teachers continue to rely solely on traditional content-driven lectures, surface learning assessment, and unconstructive feedback, they risk becoming obsolete, as mentioned in the findings. Teachers are often resistant to change and some may view AI technologies as a threat to their role as educators. A visiting professor in signals, systems, and cybersecurity featured in Times Higher Education (Farrell, 2023) also warns, "*AI will replace academics unless our teaching challenges students*," adding that "*delivery of educational material chunked at the optimal grade for retention by passive student-consumers is ripe for automation.*"

**Conclusions**

In 2008, renowned educational researcher Prof. John Hattie discovered that teachers had the most significant in-school impact on student learning, as discussed in his book "Visible Learning" (Hattie, 2008). Fifteen years later, his recently published book "Visible Learning: The Sequel" (2023) reaffirms that teachers continue to be the most influential factor in student learning success, particularly in regard to what teachers think. This remains true despite the challenges of COVID-19 and the resulting distance learning. Although in general the findings of this study follow Hattie's belief and finding with human teachers being irreplaceable. However, there is potential for these AI technologies to eventually replace teachers. As AI advancements progress, generative models' capabilities could surpass human educators' expertise and skills in various aspects of teaching and learning. This paradigm shift could lead to a reimagining of the educational landscape, with generative AI assuming a more prominent role and ultimately displacing traditional teaching roles.

In my opinion, this study is a rude awakening for teachers and universities, on one hand, our immediate concern is subsided as the findings in this study demonstrated the irreplaceable role of teachers, but on the other hand, teachers and universities need to reconsider the importance of education – what should we teach? How should we teach? Why are we teaching that with that approach? What do we really want our students to learn and develop? All these questions need to partner with technologies in mind, that may mean upskilling our teaching skills depending on the evolution of AI technologies and the values society places on human qualities in the educational process.

Table 1 highlights the unique qualities of human teachers across eight categories and 26 aspects, emphasizing their strengths compared to the limitations of AI in education. Understanding these strengths can help teachers, students, and universities make informed decisions about the future of education in a world increasingly influenced by generative AI.

For teachers, this table serves as a roadmap for areas where they can focus on refining their skills and abilities, particularly those that are difficult for AI to replicate. Emphasizing emotional intelligence, pedagogical skills, and personalized support, teachers can ensure that they remain indispensable in the educational process. Additionally, continuous professional development will allow teachers to stay ahead of advancements in AI and integrate technology effectively in their classrooms.

Students can benefit from this table by recognizing the importance of the human touch in their education. While AI can provide resources and support, the emotional and interpersonal skills of human teachers are essential for personal growth, resilience, and critical thinking. Understanding the value of human teachers will encourage students to seek out personal connections and make the most of the learning opportunities that AI cannot fully replicate.

Universities should use this table as a guideline to design curriculums that capitalize on the strengths of human teachers while leveraging AI to enhance learning, obviously costs, workload and timing need to be all accounted for. By incorporating AI as a tool to support, rather than replace, human teachers, universities can provide a comprehensive and well-rounded educational experience for their students.

As generative AI continues to emerge rapidly in education, it is crucial to focus on a symbiotic relationship between human teachers and AI. Teachers can adopt AI to handle mundane tasks, freeing up time to focus on the aspects of teaching that require a personal touch. At the same time, educators should advocate for ethical considerations in AI development, ensuring that AI systems are designed to complement, not replace, their human counterparts.

In conclusion, the future of education lies in the synergy between human teachers and AI. By understanding the unique qualities of human teachers outlined in the table, teachers can hone their irreplaceable skills, students can appreciate the value of human connection, and universities can create educational environments that effectively balance the strengths of teachers and AI.

**Limitations**

This research study comes with certain limitations, such as a comparatively small sample size that might not accurately represent all post-secondary educational institutions. Moreover, the investigation concentrated solely on text-based generative AI technology, without considering other forms or variations. Lastly, the study depended on self-reported data from participants, which could potentially introduce bias or inaccuracies.

**Declarations:**

Availability of data and material: The datasets used and/or analysed during the current study are available from the corresponding author on reasonable request

We declare no competing interests.

Acknowledgements: The author wishes to thank the students and teachers who participated the survey.